\documentclass[preprint,preprintnumbers,amsmath,amssymb]{revtex4}
\usepackage{graphicx}
\usepackage{dcolumn}
\usepackage{bm}

\begin{document}
\title{Twist versus Nonlinear Stacking in Short DNA Molecules}

\author{Marco Zoli}

\affiliation{School of Science and Technology - CNISM \\  University of Camerino, I-62032 Camerino, Italy \\ marco.zoli@unicam.it}

\date{\today}

\begin{abstract}
The denaturation of the double helix is a template for fundamental biological functions such as replication and transcription involving the formation of local fluctuational openings. The denaturation transition is studied for heterogeneous short sequences of DNA, i.e. $\sim 100$ base pairs, in the framework of a mesoscopic Hamiltonian model which accounts for the helicoidal geometry of the molecule. The theoretical background for the application of the path integral formalism to predictive analysis of the molecule thermodynamical properties is discussed. The base pair displacements with respect to the ground state are treated as paths whose temperature dependent amplitudes are governed by the thermal wavelength. The ensemble of base pairs paths is selected, at any temperature, consistently with both the model potential and the second law of thermodynamics. The partition function incorporates the effects of the base pair thermal fluctuations which become stronger close to the denaturation.  The transition appears as a gradual phenomenon starting from the molecule segments rich in adenine-thymine base pairs. Computing the equilibrium thermodynamics, we focus on the interplay between twisting of the complementary strands around the molecule axis and nonlinear stacking potential: it is shown that the latter affects the melting profiles only if the rotational degrees of freedom are included in the Hamiltonian. The use of ladder Hamiltonian models for the DNA complementary strands in the pre-melting regime is questioned.

\vskip 0.5cm

\textbf{Keywords}: DNA denaturation, Path Integral Method, Base Pair Fluctuations, Twisting of DNA Strands, Nonlinear potentials.

\end{abstract}

\maketitle

\section*{1. Introduction}
DNA is a long polymer made of millions (or even hundreds of millions) of nucleotides arranged in two complementary strands forming a double helix. Each nucleotide consists of a phosphate group and a sugar ring  disposed along the molecule backbone \emph{plus} a base, either a purine or a pyrimidine, which is essentially perpendicular to the molecule axis \cite{calla}. While phosphodiester bonds connect adjacent sugars thus giving a direction to the polynucleotide chain, bases connect the two strands via hydrogen bonds according to the Watson-Crick pairing scheme \cite{watson}. The helicoidal geometry allows (hydrophobic) bases along the strand to stack on top of each other and hide in the core of the molecule whereas (hydrophilic) phosphates and sugars form the outer shell. This is a key property of the molecule as DNA in cells is immersed in water. A ladder configuration would also allow hydrogen bond inter-strand pairing but could not allow intra-strand bases to stack on top of each other (due to structural constraints) thus leaving a hole that the surrounding water would promptly fill.

Furthermore, DNA molecules are highly compacted within a (chromosome) size which is $\sim 10^4$ times smaller than their lengths. Supercoiled DNA is in fact a semi-flexible polymer \cite{volo,kame} with a high bending rigidity due to the wrapping of the strands. Its persistence length of $\sim 500 \,$ {\AA} (in physiological ambient conditions \cite{volo11})  is:  \emph{i)} larger than its diameter ($\sim 20 \,$ {\AA}),  \emph{ii)} much larger than the base pair thickness ($\sim 3.4 \,$ {\AA}),  \emph{iii)} generally much smaller than the total molecular length scale,  as estimated through force-extension experiments on single molecules \cite{strick} first carried out by Smith et al. \cite{smith,busta}. At low applied forces, DNA behaves as an elastic polymer shaped by conformational entropic forces which oppose to the stretching \cite{gennes}; at moderate forces, the DNA contour length elongates due to structural rearrangements of the nucleotides on the chain \cite{odi,marko2} which eventually manifest in a sharp transition to the overstretching regime for high forces, say $> 65 \,pN$ \cite{gross}. While applied tensions may  interfere with the twisting of the double helix thus affecting the supercoiling degree \cite{gore,marko3}, the sharpness of the \emph{overstretching transition} signals the highly cooperative character of the base pairs (\emph{bps}) interactions which, in general, can extend over a long range along the stack.

Cooperativity is also at the heart of the \emph{melting transition}, the thermally driven separation of the two strands, which has been a focus of interest for the biophysicists community since the first observations in the mid fifties of the last century \cite{thomas,doty}. The helix-coil transition is made evident by the large increase in the UV absorption spectra due to the rearrangement of the $\pi$ electrons of the bases once the double helical stacking decreases. The percentage increase in light absorption at $\sim 260$ {nm} is proportional to the relative presence, in the heterogeneous  molecule, of adenine-thymine (AT) \emph{bps} with respect to guanine-cytosine(GC) \emph{bps} 
\cite{note1}. In fact the latter present three hydrogen bonds while the former, with two hydrogen bonds, can be more easily disrupted by thermal effects. Note that AT-\emph{bps} effective bond energies may be $\sim 30 meV$, that is just above $k_BT$ ($k_B$ is the Boltzmann constant) at room temperature. Then, whereas the \emph{overstretching transition} essentially samples the strong covalent bonds along the molecule backbone, the \emph{melting transition} results from the subtle interplay of intra-strand and inter-strand interactions providing both stability and flexibility to the molecule \cite{li}.

Besides offering considerable theoretical challenges for the statistical physics, the melting (or denaturation) transition is a template for fundamental biological functions as replication and transcription in which the double helix has to unzip, or open temporarily, to expose the bases hidden in its core and allow for a reading of the genetic code. The breathing of DNA and bubble formation  are then dynamical processes \cite{hwa,ares1,ares2,metz10,segal,handoko} crucially shaped by the strong thermal fluctuations of the \emph{bps} and also by the ambient conditions, i.e. the salt concentration in the solvent \cite{santa,owc}: the higher the density of positive ions in solutions the more screened is the electrostatic repulsion between the negatively charged complementary strands hence, the melting temperature is shifted upwards.

The first quantitative studies of DNA were based on Ising like models for which each base pair along the molecular chain is either open or closed \cite{poland,fisher}. While refined methods based on the Poland algorithm and polymer theory \cite{poland1,jost} have been later applied to compute the melting profiles, the character of the melting transition whether first  \cite{peliti,stella,carlon} or second  \cite{hanke} order remains however matter of debate \cite{seno}. Beyond the two state models, Hamiltonian approaches to denaturation based on the Peyrard-Bishop (PB) model (and its extensions) have been proposed \cite{pey1,joy05}. These models treat the base pair displacements as continuous variables  thus allowing a description of those intermediate states, essential to the DNA dynamics, that are becoming available through sophisticated experimental techniques \cite{bonnet,zocchi2,singh}. It has been recognized that local openings start in AT rich regions and eventually propagate even to portions of the helix away from the initial bubble suggesting that nonlocal effects are related to the base pair thermal fluctuations.  Thus cooperativity follows from the nonlinearities, amplified at increasing temperature, albeit peculiar of the microscopic interactions along the strand and between the strands.

On the base of these remarkable properties of the molecule, I have recently developed a computational method based on the path integral formalism \cite{io09,io10}, suitable for application to Hamiltonian models, incorporating the sequence specificity of the double helix.  It is understood that fully atomistic descriptions are computationally intractable due to the myriad of degrees of freedom even in short fragments. Accordingly theoretical investigations start from mesoscopic models representing the interactions at the level of the nucleotide units. As a main advantage, the path integral method permits to include in the calculation a large ensemble of molecule configurations with temperature dependent base pairs fluctuations. The latter are particularly strong in the short sequences  which are the focus of the present discussion. Here, after reviewing the model, I apply the method to a mesoscopic Hamiltonian accounting for the helicoidal geometry and for the main interactions at play in the helix.
The thermodynamical properties are computed on the base of an entropic approach which selects at any temperature the ensemble of helicoidal configurations contributing to the partition function. The second law of thermodynamics is thus the fundamental principle driving the description of the microscopic interactions in the path integral method.

Evaluating the melting profiles, it is pointed out that the twisting of the complementary strands around the molecule axis is a key property of the Hamiltonian. In particular, it is shown that only a twisted geometry permits to weigh those nonlinearities in the stacking potential which ultimately have an overall stabilizing role for the molecule.  While the relevance of the rotational degrees of freedom may seem intuitive, we underline that studies based on the PB Hamiltonian over the last 25 years have generally \cite{note2} represented the DNA complementary strands by a ladder model thus neglecting the twisting even in the pre-melting regime. This work intends to fill the gap in the theoretical modeling.

The Hamiltonian is discussed in Section 2 while the method and the computational techniques are given in Section 3. Some results are shown in Section 4 while Section 5 contains the final remarks.

\section*{2. Model}

In a fundamental work Englander et al. \cite{engl} suggested that the transient opening of adjacent base pairs, associated with torsional oscillations around the helix axis, could generate coherent thermally activated soliton excitations propagating along the DNA backbone. These observations have fostered a rich line of research on the nonlinear dynamics of DNA which continues nowadays \cite{sale,yakus,dani}.
Following the observations by Prohofsky \cite{proh} on the strong nonlinearities in the hydrogen bonds stretching modes, Peyrard and Bishop \cite{pey1} proposed a minimal model for two homogeneous chains, each with $N$ bases, connected by a Morse potential:

\begin{eqnarray}
& & H^{PB} =\, \sum_{n=1}^N \biggl[ \frac{\mu \dot{y}_{n}^2}{2}  +  V_S(y_n, y_{n-1}) + V_M(y_n)  \biggr] \, \nonumber
\\
& & V_S^{PB}(y_n, y_{n-1})=\, \frac{K}{2} ( y_n -  y_{n-1} )^2 \, \nonumber
\\
& & V_M^{PB}(y_n) =\, D \bigl(\exp(-a y_n) - 1 \bigr)^2  \,
\,
\label{eq:0}
\end{eqnarray}

where $\mu$ is the reduced mass of a base pair and $y_n$ is the out of phase displacement which stretches the hydrogen bond for the \emph{n-th} base pair. It measures
the pair mates separation with respect to the ground state position,  $y_n \sim 0$.   The transverse stretchings are generally much larger than the longitudinal base pairs displacements (along the molecule backbone) which are accordingly dropped. Thus the model is one dimensional. The intra-strand stacking potential is modulated by the harmonic coupling constant $K$.
$D$ is the pair dissociation energy and $a$ is the inverse length setting the range of the Morse potential. While the latter is analytically convenient, any other potential with a
hard core accounting for the phosphate groups repulsion, a stable minimum and a dissociation plateau would
be physically suitable.
The thermodynamic properties of Eq.~(\ref{eq:0}) can be described in the transfer integral method as done by Krumhansl and Schrieffer \cite{krum} in a seminal work  on the $\phi^4$ model. In the limit of a large system, $N \rightarrow \infty$, the partition function is solved exactly using the eigenvalues $\epsilon_i$ and the complete set of eigenvectors $\phi_i(y_{n})$ of the transfer integral equation

\begin{eqnarray}
& &\int d y_{n-1}\exp\biggl[-\beta \bigl(V_S(y_n, y_{n-1}) +  V_M(y_n) \bigr) \biggr] \cdot \phi_i(y_{n-1})=\,\exp(-\beta \epsilon_i)\phi_i(y_{n})  \,  \nonumber
\\
& &\delta(y_N - y_0)=\, \sum_i \phi_i^\ast(y_{N}) \phi_i(y_{0})\,,
\label{eq:32}
\end{eqnarray}

with $\beta=\,(k_B T)^{-1}$.  In the continuum regime defined by the dimensionless ratio $R \equiv \, Da^2 /K \ll 1 $,  Eq.~(\ref{eq:32}) can be approximated by the second order differential equation:

\begin{eqnarray}
& &\phi_i''(y) + 2K\beta^2 [E_i - U(y)]\phi_i(y)=\,0 \,  \nonumber
\\
& &U(y)=\,D \bigl[\exp(-2ay) - 2\exp(-ay)\bigr] \,  \nonumber
\\
& &E_i=\,\epsilon_i + \frac{1}{2 \beta} \ln \Bigl({ \frac{2 \pi}{\beta K}}\Bigr) - D \,
\label{eq:10}
\end{eqnarray}

which coincides with the Schr\"{o}dinger equation for a particle in the Morse potential after replacing

\begin{eqnarray}
K\beta^2 \rightarrow \frac{\mu}{\hbar^2}\,
\label{eq:10a}
\end{eqnarray}

where $\hbar$ is the reduced Planck constant. Thus the one dimensional statistical mechanics of the model in Eq.~(\ref{eq:0}) is mapped onto the quantum mechanics of the Morse oscillator \cite{morse}. In the latter, if a particle has mass lighter than a critical mass, it cannot be confined in the asymmetric potential well as quantum fluctuations will drive it out.

The spectrum of bound states for Eq.~(\ref{eq:10}) is given by

\begin{eqnarray}
& &E_i =\, -D \Bigl[1 - \frac{(i + 1/2)}{\delta }   \Bigr]^2\,\hskip 1cm   i=\,0, 1, ...,I(\delta -1/2) \nonumber
\\
& &\delta =\, \frac{\beta }{a} {\sqrt{2 K D}}
\label{eq:15}
\end{eqnarray}

$I(\delta -1/2)$ being the integer part of $\delta - 1/2$. It follows that, if $\delta > 1/2$, at least the ground state is bound. For $i =\,0$ and $\delta_c =\, 1/2$, Eq.~(\ref{eq:15}) defines a critical temperature $T_c$ above which all states have gone into the continuum:

\begin{eqnarray}
T_c =\, \frac{2}{k_B a} {\sqrt{2 K D}}
\label{eq:16}
\end{eqnarray}

The melting of the double helix is then formally equivalent to the disappearance of the last bound state. In the thermodynamic limit, the helix-coil transition is second-order as it is the free energy second derivative which diverges at $T_c$.

Note that general arguments regarding the impossibility of phase transitions in one dimensional systems are not applicable to Eq.~(\ref{eq:0}): \emph{i)} van Hove's theorem \cite{hove} holds for models with short range pair interactions whereas Eq.~(\ref{eq:0}) contains the unbound on site potential $V_M^{PB}(y_n)$ acting as an external field; \emph{ii)} Landau' theorem \cite{landau} states that phase coexistence cannot occur at finite temperatures as the energetic cost of making a domain wall between two regions is finite.  In fact, in the continuum limit, Eq.~(\ref{eq:0}) admits a domain wall solution connecting open and closed parts of the molecule but the domain wall energy is infinite for a large system.

As suggested by Azbel \cite{azbel1} in the context of the Ising model for heterogeneous polymers,  local strands separation releases an amount of winding entropy which ultimately drives the helix-coil phase transition. Following this idea, Dauxois-Peyrard-Bishop (DPB) \cite{pey2} have improved the model in Eq.~(\ref{eq:0}) adding nonlinearities also to the stacking potential $V_S^{PB}(y_n, y_{n-1})$ via the replacement

\begin{eqnarray}
K \rightarrow  K \Bigl[ 1 + \rho \exp\bigl[-\alpha(y_n + y_{n-1})\bigr] \Bigr]  \,.
\label{eq:17}
\end{eqnarray}

This choice is microscopically consistent with the observation that, whenever either one of the {\it bps} in Eq.~(\ref{eq:17}) is stretched over a distance larger than $\alpha^{-1}$, the hydrogen bond breaks and the electronic distribution around the two pair mates is modified. Accordingly, the stacking coupling between neighboring bases  drops from $\sim K(1 + \rho)$ to $\sim K$. Then, also the next base pair tends to open as both pair mates are less closely packed along their respective strands. Here is the link between stacking anharmonicity and {\it cooperative} effects which underlines the formation of a region with open {\it bps}.
As a base pair near an open site has lower stretching mode frequencies, its contribution to the free energy is smaller. In this context it has been proposed that the denaturation is a sharp, first order like, transition driven by the entropic gain associated to the reduced stacking coupling which renders the bases motion more disordered \cite{cule}. While the precise value of the melting entropy may depend on the sequence specificities \cite{blake}, recent neutron scattering investigations have suggested that  DNA maintains the helical order almost up to the melting \cite{eijck}  and that denaturation is rather a continuous transition, albeit occurring in a narrow temperature range as shown by renormalization group analysis of the DPB model \cite{santos}. Similar trends have been found in simulations of DNA denaturation  based on Langevin dynamics \cite{druk} and in the path integral method developed both for homogeneous \cite{io09} and heterogeneous \cite{io10} sequences, although in the latter the denaturation takes place in multiple steps due to the fact that the AT rich regions generally open at lower temperatures than the GC rich ones \cite{wart}.
While the classification of the order of the denaturation transition has been thrilling many statistical physicists for long, we let the matter open and rather focus in this paper on the interplay between stacking interactions, helix stability and torsional degrees of freedom which are essential to transcriptional initiation, requiring formation of fluctuational bubbles at specific sites \cite{benham,benham1}.

\begin{figure}
\includegraphics[height=8.0cm,width=10.5cm,angle=0]{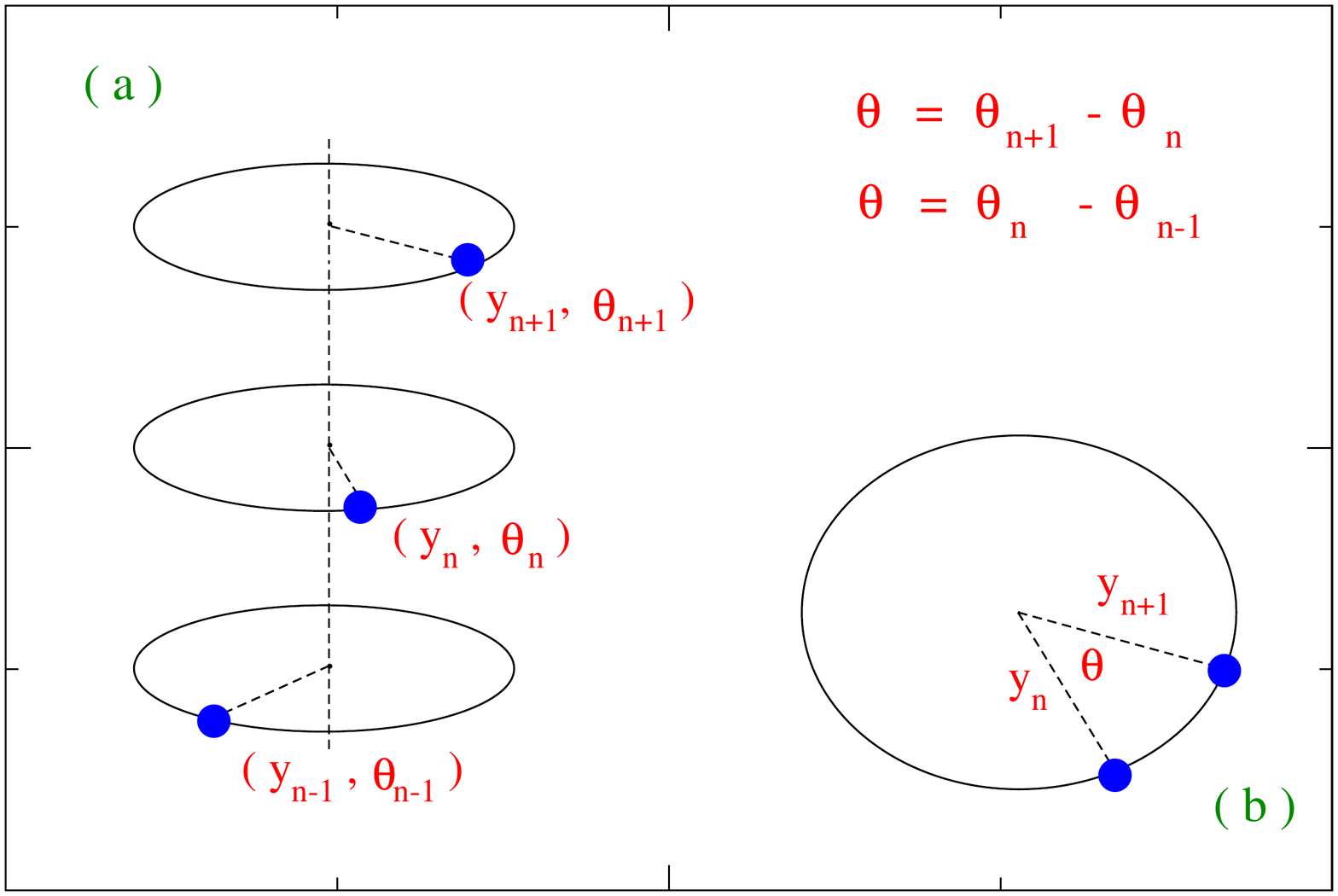}
\caption{\label{fig:1}(Color online) (a) Fixed planes picture for the right-handed helicoidal model. The blue filled circles denote the pointlike base pairs stacked along the molecule axis with twist $\theta$. The radial coordinate $y_n$ describes the $n-th$ base pair displacement from the ground state. The dashed vertical axis corresponds to the $y_n \equiv 0$ configuration, the minimum for the one-coordinate potential $V_M(y_n) + V_{sol}(y_n)$ in Eq.~(\ref{eq:1}). (b) The helix plane seen from above.}
\end{figure}

This is done in the framework of a generalized DPB Hamiltonian \cite{io11} which also includes: \emph{a)} a (positive) twist angle $\theta$ between adjacent \emph{bps} along the stack; \emph{b)} a solvent potential, whose parameters are related to the salt concentration, having the effect to enhance the Morse threshold for pair dissociation. This Hamiltonian had been previously introduced in ref.\cite{weber} where however $\theta$ takes a small, fixed value with the only aim to avoid numerical divergences in the partition function.

The model is depicted in Fig.~\ref{fig:1} and its analytical form reads:

\begin{eqnarray}
& & H =\, \sum_{n=1}^N \biggl[ \frac{\mu \dot{y}_{n}^2}{2} +  V_S(y_n, y_{n-1}) + V_M(y_n) + V_{sol}(y_n) \biggr] \, \nonumber
\\
& & V_S(y_n, y_{n-1})=\, \frac{K}{2} \Bigl[ 1 + \rho \exp\bigl[-\alpha(y_n + y_{n-1})\bigr] \Bigr] ( y_n^2 - 2 y_n y_{n-1}\cos\theta + y_{n-1}^2 ) \, \nonumber
\\
& & V_M(y_n) =\, D_n \bigl(\exp(-a_n y_n) - 1 \bigr)^2  \,
\, \nonumber
\\
& & V_{sol}(y_n) =\, - D_n f_s \bigl(\tanh(y_n/ l_s) - 1 \bigr)  \, .
\label{eq:1}
\end{eqnarray}

The sequence heterogeneity is accounted for by the site dependent dissociation energy $D_n$ and inverse length $a_n$ whereas the stacking potential parameters (harmonic $K$, anharmonic $\rho$ and $\alpha$) are taken independent of the type of base at the $n$ and $n-1$ sites.
The homogeneity assumption for the stacking relies on the observation that both types of \emph{bps} contain a purine plus a pyrimidine, the former being larger and heavier. Thus the AT- and GC- \emph{bps} are comparable in size and weight.

Note that the Morse potential inflection point lies at $y_n^*=\,\ln 2/a_n$. At such point the potential opposes the maximum force ($=\,D_n a_n/2$) to the breaking of the base pair bond. This offers some criterion to estimate the potential parameters \cite{zdrav} also at the light of DNA mechanical opening experiments \cite{heslot} which however sample the breathing of groups of \emph{bps} along the helix.
Usually assumed effective values for the pair dissociation energies are
$D_{AT}=\,30 meV$ and $D_{GC}=\,45 meV$ while  $a_{AT}=\,4.2$ {\AA}$^{-1}$ and $a_{GC}=\,5$ {\AA}$^{-1}$ in view of the fact that AT-base pairs have larger displacements than GC-base pairs \cite{zhang}.
Larger dissociation energy values \cite{campa} would shift the denaturation steps at higher $T$ as the melting temperatures of synthetic homopolymers vary essentially linearly with $D_n$ (instead, Eq.~(\ref{eq:16}) holds in the strong $K$ limit).

As for the stacking, the commonly used parameters are: $K=\, 60meV$ {\AA}$^{-2}$,  $\rho=\,2$,  $\alpha=\,0.5$ {\AA}$^{-1}$ whereas the anharmonicity effects due to larger $\rho$ are discussed below.  Somewhat smaller (or larger) $K$  i.e.: $K \sim \, 25meV$ {\AA}$^{-2}$ \cite{campa}, $K \sim \, 1eV$ {\AA}$^{-2}$ \cite{barbi0},  would have the main effect to shift downwards (or upwards) the denaturation temperature (see Eq.~(\ref{eq:16})) without changing the character of the transition nor the shape of the melting profiles \cite{io11a}. Taking the ratio $\alpha / a_n \ll 1$, one weighs the stacking nonlinearities assuming that the range of the fluctuations in $V_S$  is larger than that of $V_M$.
Incidentally, we note that a recent parametrization of the (harmonic) PB model parameters has been carried out also for RNA by fitting the melting temperature data \cite{weber1}.

The solvent term $V_{sol}$ adds to $V_{M}$ thus enhancing by  $f_s D_n$ the height of the energy barrier above which the base pair dissociates \cite{collins,singh1,pablo}.
Taking a factor $f_s=\,0.3$, one assumes a high salt concentration  which screens the negatively charged phosphate groups \cite{large}.
The length $l_s=\,3 ${\AA} defines the range beyond which the Morse plateau is recovered and $D_n$ returns to be the fundamental energy scale.
For  $y_n > l_s$, the two strands are apart from each other and the hydrogen bond with the solvent is established.

The angle $\theta$ in Eq.~(\ref{eq:1}) is an input parameter of the model: the twist can be tuned but once it is assigned, it is kept constant throughout the whole investigated temperature range. This assumption has important consequences. In fact, the presence of a finite $\theta$  produces a local term in the stacking $V_S$ of Eq.~(\ref{eq:1}) which competes on the energy scale with the plateau of $V_M$. If $\theta$ remains sufficiently large also in the high $T$ range, $V_S$ has a stabilizing effect on the molecule as the torsion provides some restoring force which opposes to the strands separation \cite{io12}. This amounts to say that, as long as an helix exists, the exact phase transition peculiar of the DPB model with zero twist cannot occur.

However experiments have shown since long that the average helix rotation angle decreases by increasing $T$ \cite{depew,duguet}.  
Therefore, in a consistent analysis of the thermally driven helix unwinding, $\theta$ should be a ($T$ dependent) output of the computation and, in the high $T$ limit, one should get, $\theta \rightarrow 0$. Hence, the ladder model of the denaturated phase should be recovered. With this caveat, keeping the twist as a tunable parameter, we can still evaluate the effects of the helicoidal geometry on the melting profiles being aware that, more precisely, we are sampling a pre-melting regime in which a complete strand separation has not yet been achieved. This is done after discussing the theoretical background of the path integral method.

\section*{3. Path Integral Method}

In the path integral approach to quantum mechanics, the transition amplitude $< x_f, t_f | x_i, t_i>$ of the time evolution operator between two localized particle states is a sum over all histories (paths) along which the particle may evolve in going from the initial state $| x_i, t_i>$ to the final state $| x_f, t_f>$ \cite{feyn}. The quantum mechanical amplitude is used to obtain the quantum mechanical partition function which, in the configuration space and in the continuum limit, reads:

\begin{eqnarray}
& &Z_{QM}=\, \oint {D}x \exp \bigl(i {A[x]}/\hbar \bigr) \, \nonumber
\\
& &{A[x]}=\, \int_{t_i}^{t_f} dt L\bigl(\dot{x}(t), x(t)\bigr)\, .
\label{eq:18}
\end{eqnarray}

$\oint D x$ is the measure of integration over all paths having equal end points, $x_f(t_f)=\, x_i(t_i)$, while
$L\bigl( \dot{x}(t), x(t)\bigr)$  is the Lagrangian taken along a specific path, i.e. \emph{the difference} between kinetic and potential energy. In the $\hbar \rightarrow 0$ limit, the path satisfying the Newton's law is that for which the action ${A[x]}$ is the least \cite{feynman}.
At finite temperature, the equilibrium properties of a system are determined from the quantum statistical partition function after performing an analytic continuation of $Z_{QM}$ to the imaginary time $\tau=\,it$:

\begin{eqnarray}
& &Z=\, \oint {D}x \exp \bigl( - {A_e[x]} / \hbar \bigr) \, \nonumber
\\
& &{A_e[x]}=\, \int_{0}^{\hbar \beta} d \tau \Bigl( \frac{\mu }{2} \dot{x}^2(\tau) + V( x(\tau), \tau) \Bigr)\, \nonumber
\\
& &\hbar \beta \equiv i(t_f - t_i) \, .
\label{eq:19}
\end{eqnarray}

Accordingly, in the statistical formalism, the Euclidean action ${A_e[x]}$ replaces the mechanical canonical action ${A[x]}$ and the partition function is an integral in the configuration space over paths running along an imaginary time axis. Note the disappearance of the complex $i$ in the path integral of Eq.~(\ref{eq:19}). Unlike Eq.~(\ref{eq:18}) there are no oscillating phases in the exponent hence, the largest contribution to $Z$ comes from those paths for which the \emph{the sum} of kinetic energy and potential energy ($V$) is small.

As the paths are closed trajectories, $x(0)=\,x(\beta)$,  they can be expanded in Fourier components

\begin{eqnarray}
& &x(\tau)=\, x_0 + \sum_{m=1}^{\infty}\Bigl[a_m \cos({2 m \pi} \tau / {\beta}) + b_m \sin({2 m \pi} \tau / {\beta}) \Bigr] \,,
\label{eq:6a}
\end{eqnarray}

therefore allowing to define the path integral measure in Eq.~(\ref{eq:19}) as:

\begin{eqnarray}
& &\oint {D}x \equiv \frac{1}{\sqrt{2}\lambda_Q} \int dx_0 \prod_{m=1}^{\infty}\Bigl(\frac{m \pi}{\lambda_Q} \Bigr)^2 \int da_m \int db_m \, \, \nonumber
\\
& &\lambda_Q= \sqrt{{\pi \beta \hbar^2} / {\mu}} \,.
\label{eq:3b}
\end{eqnarray}

Eq.~(\ref{eq:3b}) normalizes the kinetic term in ${A_e[x]}$  while $\lambda_Q$ is the thermal wavelength in the quantum case \cite{io05a,io05b}.

\subsection*{A. Path Integral for DNA }

I have applied the imaginary time path integral method  to Eq.~(\ref{eq:1}) by introducing the idea that the base pair displacements $y_n$ can be described by paths $x(\tau_i)$ as in Eq.~(\ref{eq:6a}). The index $i$ numbers the $N$ \emph{bps} along the $\tau$-axis.
In fact there are $N + 1$ base pairs in Eq.~(\ref{eq:1}) but the presence of an extra base pair $y_0$ is remedied by taking periodic boundary conditions, $y_0 = \, y_N$, which close the finite chain into a loop.  This condition is  incorporated in the path integral description as the path is a closed trajectory. Hence a molecule configuration is given by $N$ paths and, in the discrete time lattice, the separation between nearest neighbors \emph{bps} is $\Delta \tau =\,\beta / N$. Then, Eq.~(\ref{eq:1}) transforms onto the time axis by mapping: $y_n \rightarrow x(\tau_i)$ and $y_{n-1} \rightarrow x(\tau_i - \Delta \tau)$.

As DNA denaturation occurs at high temperatures the thermal wavelength in Eq.~(\ref{eq:3b}) has to be adapted to the classical case. This is done  by using Eq.~(\ref{eq:10a}) which leads to define the classical thermal wavelength as:

\begin{eqnarray}
\lambda_{C}= \sqrt{{\pi} / {\beta K}} \,.
\label{eq:3c}
\end{eqnarray}

\begin{figure}
\includegraphics[height=8.0cm,width=10.5cm,angle=0]{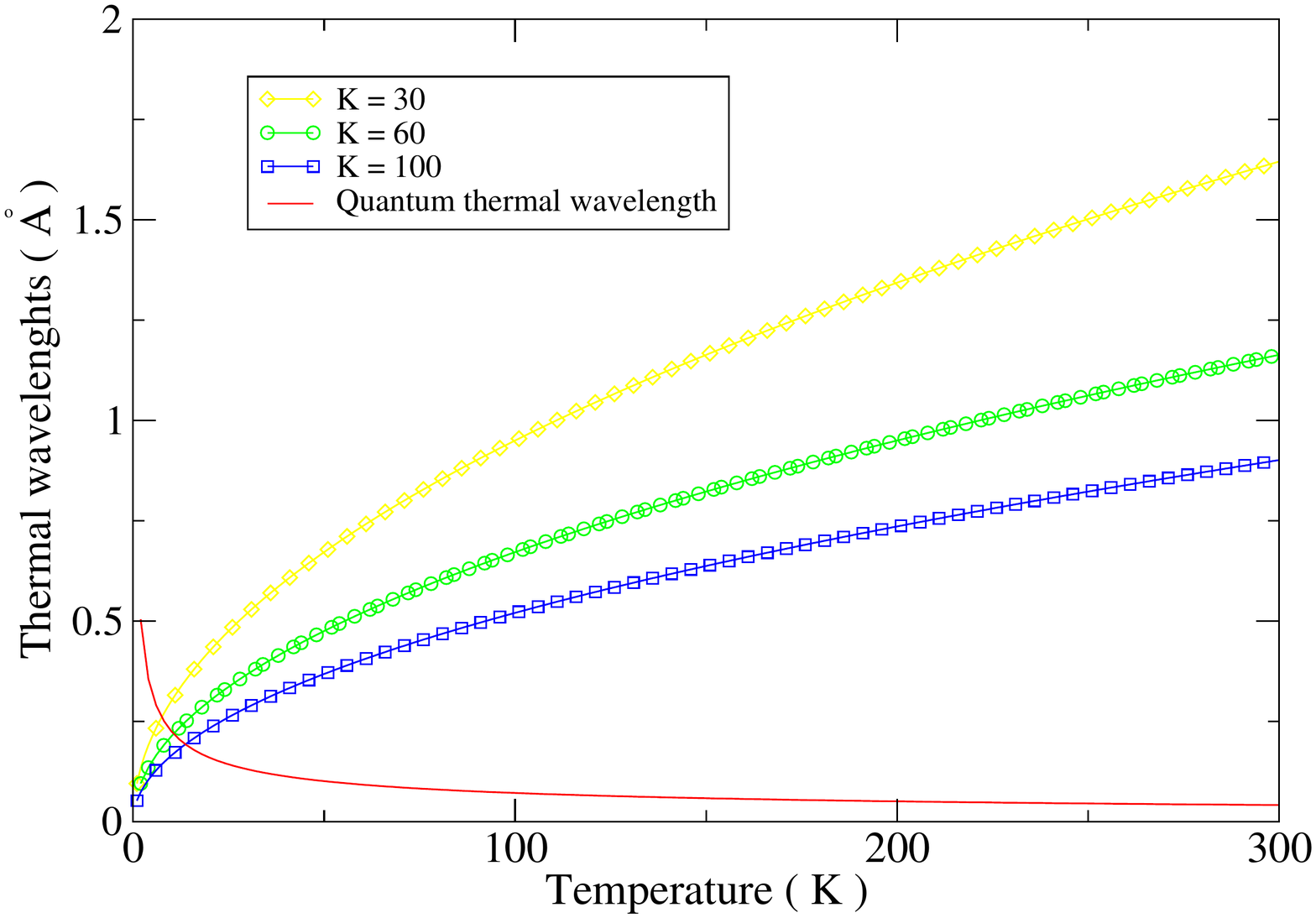}
\caption{\label{fig:2}(Color online) Quantum and classical thermal wavelengths computed from Eq.~(\ref{eq:3b}) and Eq.~(\ref{eq:3c}), respectively. The harmonic stacking couplings are in units $meV$ {\AA}$^{-2}$.}
\end{figure}

The stabilizing effect of the harmonic stacking coupling is visualized in Fig.~\ref{fig:2} where $\lambda_{C}$ is plotted  together with $\lambda_{Q}$ versus temperature. The intersections between quantum and classical curves select the low $T$ regime in which quantum effects become relevant.

Applying the space-time mapping technique, the classical partition function for the DNA Hamiltonian in Eq.~(\ref{eq:1}) reads:

\begin{eqnarray}
& &Z_C=\oint {D}x\exp\biggl[- \beta A_C[x] \biggr]\,  \, \nonumber
\\
& &A_C[x]=\,\sum_{i=\,1}^{N} \Bigl[\frac{\mu }{2} \dot{x}(\tau_i)^2 + V_S(x(\tau_i),x(\tau_i - \Delta \tau)) + V_M(x(\tau_i)) + V_{sol}(x(\tau_i)) \Bigr]\, ,\,\nonumber
\\
\label{eq:6c}
\end{eqnarray}

from which the free energy $F=\,\beta^{-1}\ln Z_C$, the entropy $S=\,- dF/dT$ and the constant volume specific heat $C_V=\, T dS/dT$ can be derived.

Computing the average path amplitudes by:

\begin{eqnarray}
< x(\tau_i) >=\,Z_C^{-1}\oint {D}x x(\tau_i) \exp\bigl[- \beta A_C\{x\}\bigr] \,, \,
\label{eq:6d}
\end{eqnarray}

one can derive the melting profiles, locate the bubble formation along the sequence and monitor the bubble size as a function of $T$.

\subsection*{B. Computational method }

As the \emph{bps} displacements are Fourier expanded, see Eq.~(\ref{eq:6a}), a set of coefficients $\{x_0 , a_m , b_m\}$ selects a point in the path configuration space $\{x(\tau_i), \, i =\,1\,,..,\,N \}$ which corresponds to a molecule state. The Fourier integrations, see Eq.~(\ref{eq:3b}), sample the path space
thus building an ensemble of distinct configurations for the system. As this is repeated for any temperature, we have a tool to describe the base pair thermal fluctuations around the equilibrium ($x(\tau_i) \sim 0$).

In principle the configurations ensemble for the DNA fragment is infinite as it may include any possible combination of Fourier coefficients. For practical purposes some criteria intervene to select computationally those good paths, defining physical molecule states, which contribute to the partition function. This poses a restriction on the ensemble size.

Such criteria are of two types:

\emph{1)} $V_M(x(\tau_i))$ may admit also negative base pair paths corresponding to a inter-strand compression with respect to the equilibrium. Too negative paths are however forbidden due to the hard core repulsion between negatively charged sugar-phosphate groups. The programme selects a lower bound for any base pair amplitude, $x_{min}(\tau_i) < 0$, such that $V_M (x_{min}(\tau_i)) + V_{sol} (x_{min}(\tau_i)) \sim D_n$ thus avoiding unphysical divergences in the partition function \cite{zhang}.
The integrals in Eq.~(\ref{eq:3b}) require a cutoff which operates a truncation in the configuration space \cite{zhang}. By normalizing the kinetic part of ${A_C[x]}$, the cutoff is found to be $\propto \lambda_C$, namely $\propto \sqrt{T}$, according to Eq.~(\ref{eq:3c}). This is an important feature in the computational method as it allows to build a $T$ dependent ensemble of path amplitudes consistently with the physical expectations.

\emph{2)} The path ensemble has to fulfill the second law of thermodynamics hence, the helical configurations are shaped by an entropic approach. First, the code computes the entropy at a given initial $T$ achieving numerical convergence by including a sufficiently high number of paths $( \sim 10^6)$. Next, the code proceeds to the successive temperature step and evaluates the entropy difference  with respect to the first value. If the entropy is \emph{not} growing versus $T$ a new partition is performed in the Fourier coefficients integration, a new path ensemble is selected and the entropy is recalculated. This is done at any $T$ until the macroscopic constraint of the second law of thermodynamics is fulfilled throughout the whole investigated temperature range. I emphasize that the method does not put any constraint on the shape of the \emph{entropy versus $T$}- plot aside from the requirement that the entropy derivative has to be positive.
It is found that the size of the path ensemble grows versus $T$ as an increasing number of paths cooperatively leads to bubble formation along segments of the double helix when the melting regime is approached.
The size of the ensemble is therefore a measure of the cooperativity of the system. In general, heterogeneous sequences display a lower degree of cooperativity than homogeneous ones as the denaturation process occurs in multisteps \cite{io10}.

\section*{4. Results }

While the path integral method can be applied in principle to any sequence, we take here
a fragment with $N_\tau =\,100$ base pairs thus choosing a kind of short sequence with strong fluctuational effects for which a broadening of the denaturation range has been predicted by different methods \cite{druk,palmeri,joy07}.

The first $\tau_{i=\,1},...,\tau_{i=\,48}$ sites from left have the same sequence as the \emph{L48AS }  fragment of Refs.\cite{zocchi2} which shows fluctuations in the AT-rich side triggering the opening of distant GC pairs. In fact the UV absorption measurements show that the base pairs open essentially in two stages, at two distinct temperatures. However, the fraction of base pairs opening in the first stage is larger than the whole fraction of AT- base pairs in the sequence. This implies that also GC- base pairs, embedded in AT-rich regions, may open in particular in the presence of TATA boxes. With respect to the \emph{L48AS} sequence, $52$ AT-base pairs are added on the right side to further develop fluctuational openings \cite{choi,rapti}.

The $\tau_{i=\,101}$ site closes the double helix into a loop and therefore hosts a GC base pair. Inside AT-rich regions, pyrimidine-purine stacking steps favor local unwinding hence TA/TA steps are more efficient than AA/TT in driving bubble formation \cite{yakov,metz06,mazza}.  The whole double strand sequence, labeled L48AT22, is:

\begin{eqnarray}
& &GC + 6AT +  GC + 13AT + 8GC + AT + 4GC + \nonumber
\\
& &AT + 4GC + AT + 8GC + [49-100]AT + GC \, .
\label{eq:6}
\end{eqnarray}

In equilibrium conditions most of DNA molecules have about ten base pairs per helical turn, $h_{eq} \sim 10$. Then the equilibrium twist angle in Eqs.~(\ref{eq:1}),~(\ref{eq:6c}) is $\theta_{eq}=\, 2\pi /h_{eq}$.
In short molecules one may neglect the spatial coiling of the molecule axis around itself and focus on the twist accounting for the coiling of the individual strands around the helical axis  \cite{bates}. The equilibrium twist number is defined as  $({Tw})_{eq} \sim N / h_{eq}$  where $\sim$ stays for the closest integer number to $N / h_{eq}$.
For the present case, $({Tw})_{eq}=\,10$. Unwinding the double helix produces configurations with $Tw < ({Tw})_{eq}$ whereas overwound molecules have $Tw > ({Tw})_{eq}$.

\subsection*{A. Melting profiles }

The melting temperature is experimentally defined as the temperature at which half of the molecules in the sample are in the double-helical state and half are in the single strand, random-coil state \cite{doty1}.
On the theoretical side, the problem arises to define when a configuration is open or closed. There is necessarily some arbitrariness intrinsic to the Hamiltonian model as this is expressed in terms of  base pair stretchings which are continuous variables. Accordingly
one may assume that a configuration is open when all base pairs are larger than a certain threshold $\zeta$ which however cannot be set univocally \cite{ares1,joy09}.
The fraction of open base pairs $F_{op}$ for the system in Eq.~(\ref{eq:6}) is computed taking different $\zeta$ and searching for a range of values which yield a description of the denaturation  qualitatively consistent with that provided by thermodynamical indicators such as the specific heat \cite{wart}. This reasoning is inspired by the observation that the fraction of closed base pairs, $1 - F_{op}$, is a measure of the system internal energy hence $d F_{op} / dT$ is proportional to the specific heat. This holds for a homogeneous chain but also in heterogeneous DNA the specific heat is an indicator of the melting \cite{bresl} as it displays sharp peaks at the temperatures where various parts of the sequence open. Thus some correlation is expected between $F_{op}$ and specific heat plots versus temperature.
As the UV signal changes quite abruptly when base pairs dissociate,   $F_{op}$ is defined in terms of the Heaviside function $\vartheta(\bullet)$ as

\begin{figure}
\includegraphics[height=7.0cm,width=8.5cm,angle=0]{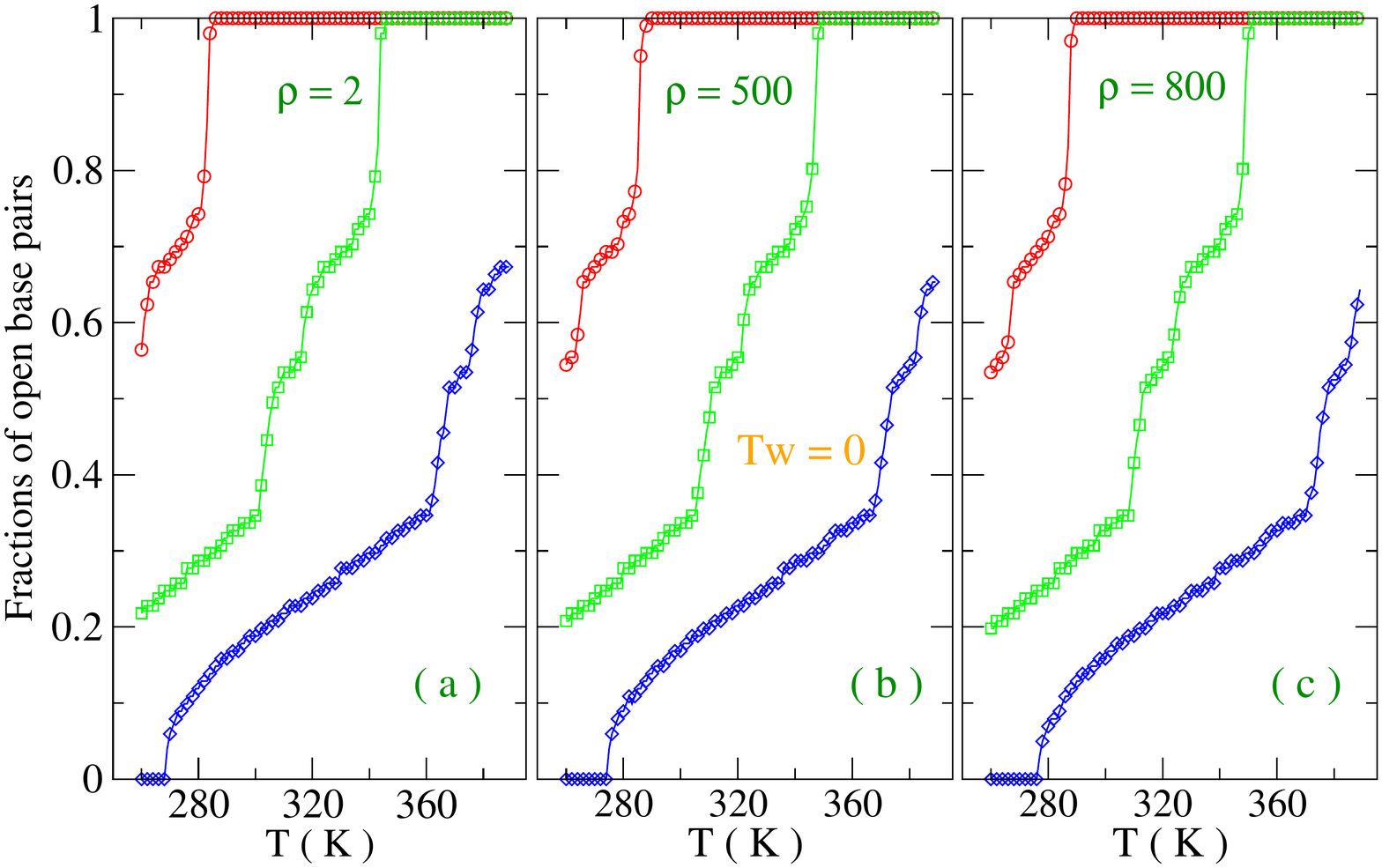}
\includegraphics[height=7.0cm,width=8.5cm,angle=0]{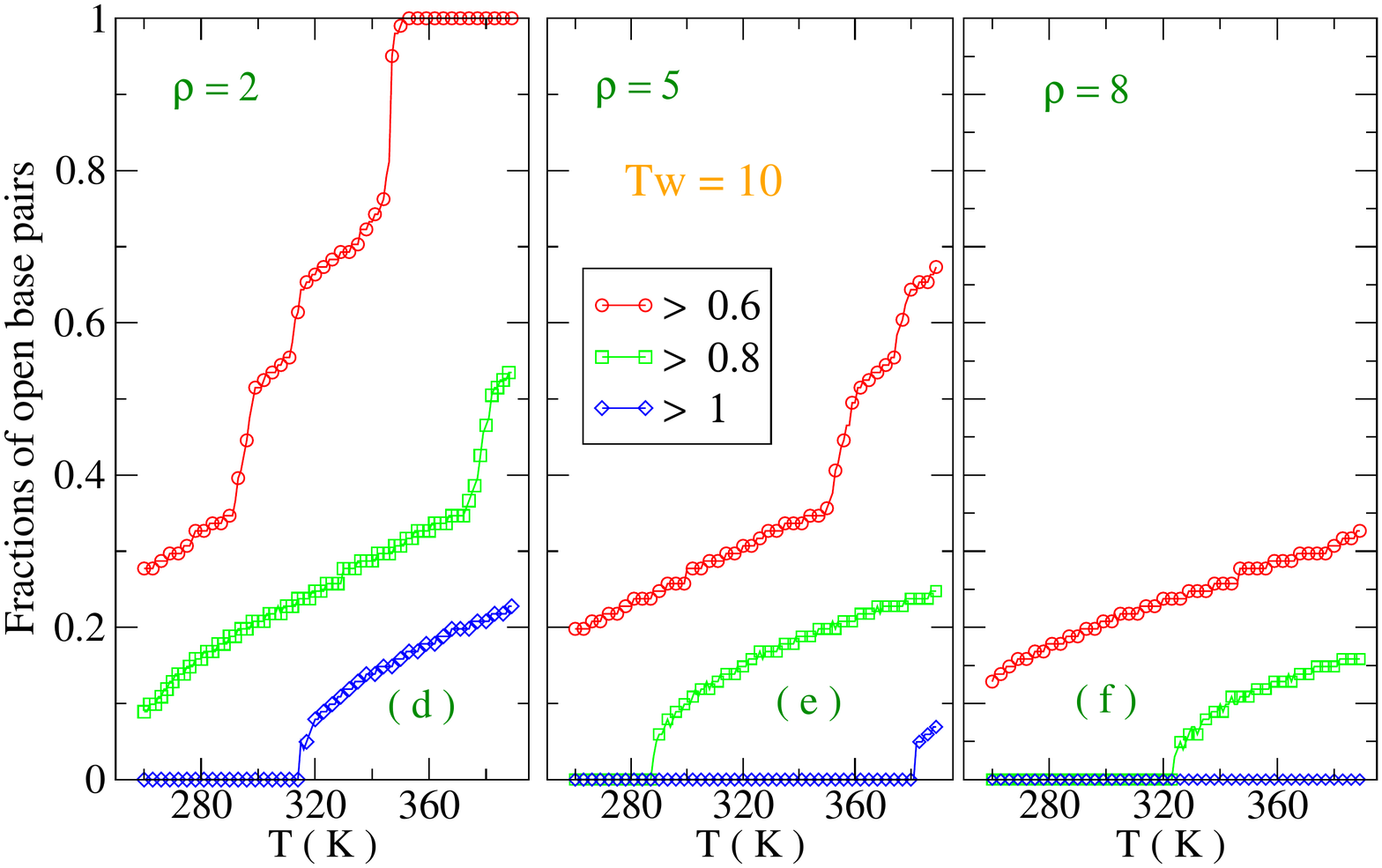}
\caption{\label{fig:3}(Color online) Fractions of average displacements larger than $\zeta=\,0.6$  (circles),  0.8 (squares),  1 {\AA} (diamonds)  versus temperature in the  DPB ladder model ((a) - (c)) and in the equilibrium twist conformation ((d) - (f)). Three anharmonic stacking $\rho$ are assumed both in the left and in the right panel. In {(e)} and {(f)}, $\rho$ is smaller by a factor $100$ than in (b) and (c) respectively. }
\end{figure}

\begin{eqnarray}
F_{op} =\, \frac{1}{N} \sum_{i=1}^{N} \vartheta\bigl(< x(\tau_i) > - \zeta \bigr) \,
\label{eq:8}
\end{eqnarray}

with $< x(\tau_i) >$ given  by Eq.~(\ref{eq:6d}) in the path integral formalism.
Then, the {\it threshold} $\zeta$ yields a criterion to establish whether an average base pair displacement is open, $<x(\tau_i)>\, \geq\, \zeta$, or not.

In Figs.~\ref{fig:3} the melting profiles are plotted \emph{both} for the DPB model with zero twist and \emph{for} the equilibrium twist conformation. The harmonic coupling is $K=\, 60 meV$ {\AA}$^{-2}$. Several $\rho$ values are assumed in both cases to investigate the interplay between twisting and stacking anharmonicity.

The $Tw=\,0$ system shows little dependence on $\rho$:  large increments (see Figs.~\ref{fig:3}(b),(c)) over the $\rho=\,2$ value (Fig.~\ref{fig:3}(a)) produce only slight variations in the denaturation patterns. Small effects are found also for $\rho < \,2$ pointing to a substantial irrelevance of the $\rho$ driven anharmonicity for the DPB ladder model.
These findings can been explained by comparing the energetics of the stacking potential to that of the hydrogen bond potential (modified by the solvent): in the absence of a twist, adjacent \emph{bps} along the stack may undergo large amplitude fluctuations with a low energetic cost. This means that, if the  \emph{n-bp} experiences a large fluctuation, the neighbor \emph{(n-1)-bp} offers little restoring force and $V_S(y_n, y_{n-1})$ in Eq.~(\ref{eq:1}) remains of order of the Morse plateau energy even for large $y_n - y_{n-1}$ \cite{io12}. Accordingly $V_S(y_n, y_{n-1})$ gives little stability to the double strand molecule as \emph{it does not sufficiently discourage large relative bps fluctuations} which therefore contribute to the partition function. This result holds both for the harmonic and anharmonic model. Accordingly our method confirms the instability of the ladder structure already envisaged by the denaturation simulations in short sequences carried out by Drukker et al. \cite{druk}.

Quite different is the physical picture emerging from the $({Tw})_{eq}$ conformation: even slight enhancements over the $\rho=\,2$ value shift upwards (along $T$) the opening of the average\emph{ bps } displacements and flatten the melting profiles suggesting that the denaturation becomes even more gradual.  As a main feature, in Fig.~\ref{fig:3}(d), the melting curve shows that: \emph{i)}  $\sim 70\%$ of the average \emph{bps} are larger than $\zeta = \, 0.6$ {\AA} at $T \sim 320K$; \emph{ii)} for the same  $\zeta$, $F_{op} \sim 1$ at $T \sim 340K$.
Note that $\rho$ values in Figs.~\ref{fig:3}(e),(f) are two orders of magnitude smaller than in Figs.~\ref{fig:3}(b),(c) respectively. The $\rho=\,8$ plot says that, even at $T \sim \,400K$, there are no average displacements larger than $1$ {\AA} while only $35\%$ are larger than $0.6$ {\AA}. Thus a relevant feature is found in the equilibrium twisted configuration by increasing $\rho$, under the assumption that $K$ is kept constant. A helicoidal geometry has the main consequence to bring closer \emph{bps} which would be distant along a chain thus favoring long range interactions \cite{yera1} and an overall stabilizing effect in the stacking \cite{cooper}.

As explained in Section 2, $\rho$ accounts in our mesoscopic model for those cooperative effects which are known to shape the molecule biological functioning at ambient temperature. The results presented in 
Figs.~\ref{fig:3} indicate that the torsion of the strands is essential to the model in order to weigh such effects and tune the stacking nonlinearities in the molecules.

Here I have discussed only the $({Tw})_{eq}$ case but the computation can be carried out both for unwound and overwound configurations with melting profiles showing a specific dependence on ${Tw}$.
For our parameter sets, significant fractions of \emph{bps} maintain displacements smaller than $1$ {\AA} even at $T \sim 370K$. This means that our molecules are not fully denaturated in this $T-$ range consistently with the fact that a helicoidal geometry is preserved. In this pre-melting regime, we observe the formation of local bubbles which are physiological to the molecule although the latter does not melt entirely. Fluctuational openings are formed as the nonlinear cooperative interactions along the molecule backbone amplify the effect of a \emph{bp} hydrogen bond disruption.

While the path integral method has been applied to the toy-sequence in Eq.~(\ref{eq:6}), the analysis may equally hold for any short sequence whose melting profiles and bubble statistics had to be experimentally known. In this case, fitting the model to the experiments at various temperatures,  one 
may deduce reliable estimates both for the anharmonic potential parameters and for the threshold $\zeta$ in Eq.~(\ref{eq:8}). Moreover, as also the melting profiles can be experimentally obtained both for unwound and overwound configurations of a given sequence, one may monitor the parameter dependence on the induced torsional stress. Here we see the applicative potential of our method with regard to the characterization of short heterogeneous sequences.

\subsection*{B. Free Energy Derivatives}

\begin{figure}
\includegraphics[height=8.0cm,width=10.5cm,angle=0]{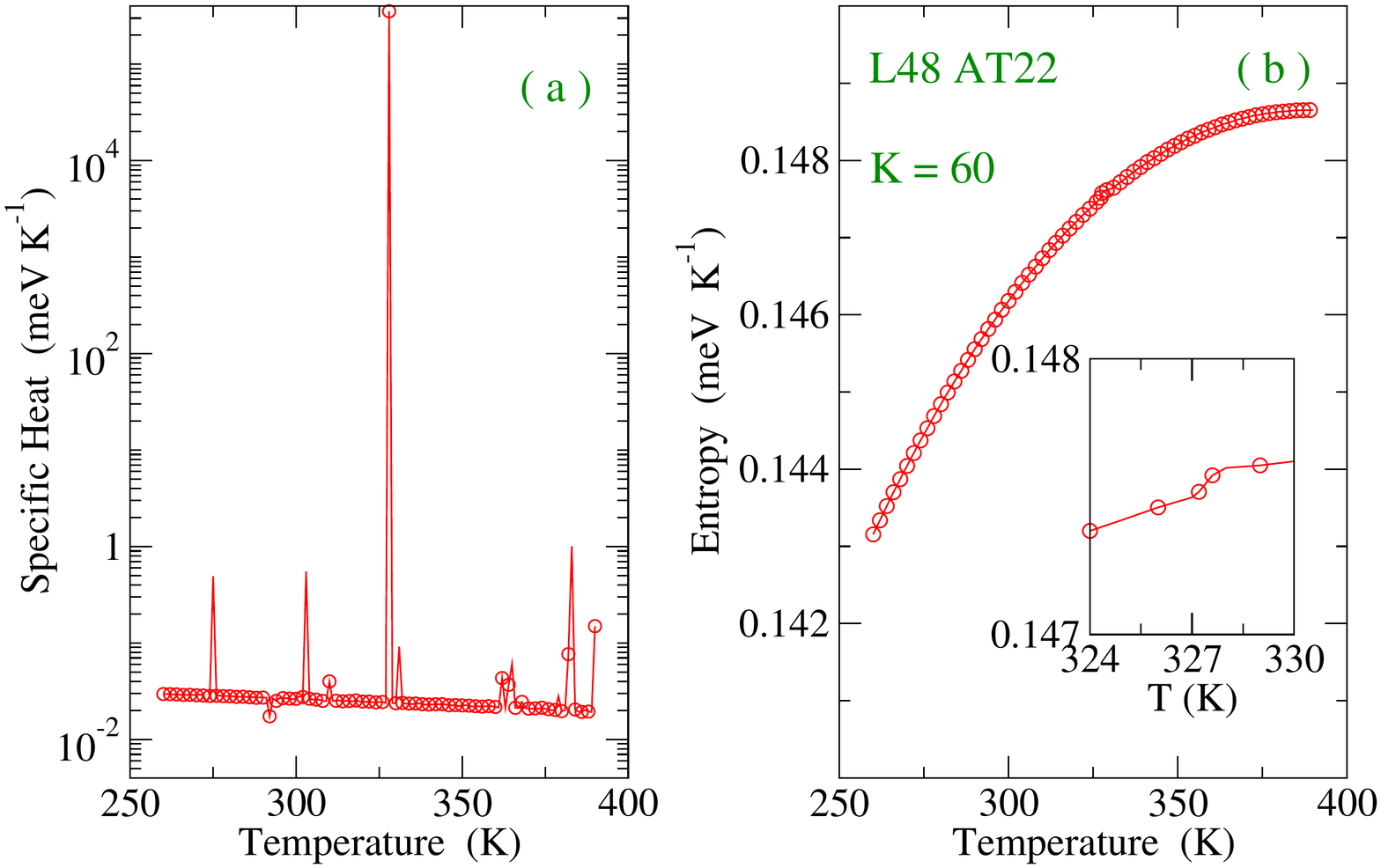}
\caption{\label{fig:4}(Color online) (a) Specific heat and (b) Entropy in the  DPB ladder model for the sequence in Eq.~(\ref{eq:6}). The solvent potential is switched off while Morse and stacking potential parameters are as in Fig.~\ref{fig:3}(a). }
\end{figure}

Fig.~\ref{fig:4} shows specific heat and entropy for the sequence in Eq.~(\ref{eq:6}). Here the solvent potential is switched off  ($f_s=\,0$) to allow a closer comparison with the standard DPB model.  Accordingly the $Tw =\,0$ conformation is assumed.  The main peak in $C_V$ at $T \sim 327K$ essentially mirrors the large increase of $F_{op}$ (see Fig.~\ref{fig:3}(a)) for $\zeta = \, 0.8$ {\AA} although $F_{op}$ grows from $\sim 0.4$ to $\sim 1$ in the broader temperature range $\sim 320 - 340K$.
This follows from the fact that the solvent tends to stabilize the system smearing the transition over a wider region. However, even without solvent potential effect, the denaturation appears as an overall smooth phenomenon in Fig.~\ref{fig:4}(b) where the entropy plot shows a slight kink at $T \sim \,327K$ (magnified in the inset).
The size of the entropy step is $\sim 10^{-4} meV K^{-1}$.  In general, our entropic gains are associated to local fluctuational openings in short sequences, hence far from the thermodynamic limit. Thus, the path integral calculation suggests an overall smooth entropy behavior in line with the transfer matrix analysis of the Joyeux-Buyukdagli model for a finite sequence of 100 base pairs  \cite{joy07}.  The latter, it is worth noticing, also estimates reduced melting entropy (with respect to the DPB ladder model) at the thermodynamic limit.

\section*{5. Conclusions }

I have discussed the theoretical background and general motivations for a path integral description of the DNA thermodynamics with particular focus on short molecules for which the sequence specificities can be probed at the level of the single base pair, the fundamental scale of the genetic code. 

An extended Peyrard-Bishop Hamiltonian accounting for the twisting of the complementary strands around the molecule backbone has been used to model the main interactions at play in the system i.e., hydrogen bonds between base pair mates and nonlinear couplings between first neighbor base pairs along the molecule stack.
The competition between the energetics of hydrogen bonds and stacking potential determines the conditions for the stability of the molecule. 

It is found that the twisting is an essential property of the mesoscopic Hamiltonian in order to weigh the effects of the nonlinear stacking in that pre-melting regime in which the molecule still preserves the double helical structure. Instead, the standard ladder Hamiltonian is inadequate in that regime. Our results are in line with recent Langevin simulations of DNA dynamics \cite{palm13} suggesting that the twist should be included in the mesoscopic model in order to fit the experimental closure lifetimes of denaturation bubbles. 

Central to our investigation are the base pair thermal fluctuations that are fully incorporated in the computation of the partition function. This is done by treating the base pair stretchings as dynamical paths whose amplitudes grow as a function of temperature. The $T$ dependence of the path amplitudes is tuned by the classical thermal wavelength (Fig.~\ref{fig:2}) in the path integration measure. Size and composition of the path ensemble are selected by the numerical code, at any temperature, imposing that the system fulfills the second law of thermodynamics. 
The method is applied to a short sequence with $N=\,100$ base pairs whose melting profiles reveal pronounced increments at various $T$ due to the successive openings of molecule portions, starting from those rich in AT- base pairs. While these features are consistent with the sharp enhancements found in the UV absorption spectra, it has to be emphasized that our small scale molecules are far from the thermodynamic limit in which a true, highly cooperative phase transition is known to occur. The body of results provided by the path integral method sees the denaturation transition as an overall smooth phenomenon mainly witnessed by the continuous (versus $T$) behavior of the entropy and taking place in a broad temperature window. This character is independent of the presence of a  potential related to the salt concentration in the solvent and seems to hold in the whole range of potential parameters, at least for the fixed planes model here assumed. The solvent potential has nonetheless a stabilizing effect on the molecule as it enhances the hydrogen bonds dissociation energy with respect to the Morse plateau.

The obtained melting profiles show that fractions of base pairs may open already at room temperature (and even below) as a result of the fluctuational effects included in our statistical method.
Large fluctuations start at single base pair sites and propagate along the molecule stack thus forming
those transcriptional openings which have biological relevance at constant, physiological temperatures. A more detailed analysis of the bubble dynamics is however required in order to detect the transcription starting sites in DNA molecules.

Eventually, we point out that the twist number in the helicoidal geometry has been treated as an input parameter of the model and the system thermodynamics has been computed keeping the twist constant in the whole $T$ range. While this approach suffices to the aims of the present study, in a more sophisticated (and CPU time consuming description), the twist should be rather an output of the numerical code to be determined on the base of a criterion of energetic convenience. Accordingly, the path integral simulation of the molecule ensemble should yield the twisted stacking configurations appropriate to the various temperatures and the helix unwinding should be signalled by specific entropic gains. 
This trend is supported by a latest application of the method \cite{io13} to a short, circular sequence recently found in mammalian cells \cite{dutta}. Further research on small molecules is on the way.
In this regard, refined mesoscopic models accounting for strong fluctuations and torsional effects may have predictive capability of the stability properties and the dynamical processes of bubbles formation in specific sequences.


\begin{thebibliography}{widest-label}

\bibitem{calla}
C.R. Calladine, H.R. Drew,   \emph{Understanding DNA}, Academic Press, San Diego, USA, 1992.

\bibitem{watson}
J.D. Watson, F.H.C. Crick,   "Molecular structure of nucleic acids," \emph{Nature} \textbf{171}, pp. 737-738, 1953.

\bibitem{volo}
A. V. Vologodskii,  "DNA Extension under the Action of an External Force," \emph{Macromolecules} \textbf{27},  pp. 5623-5625, 1994.

\bibitem{kame}
M.D. Frank-Kamenetskii,  "Biophysics of the DNA molecule," \emph{Phys. Rep.} \textbf{288}, pp. 13-60, 1997.

\bibitem{volo11}
S. Geggier,  A. Kotlyar, A. Vologodskii,  "Temperature dependence of DNA persistence length," \emph{Nucl. Acids Res.} \textbf{ 39}, pp. 1419-1426, 2011.

\bibitem{strick}
T.R. Strick, J.F. Allemand, D. Bensimon, A. Bensimon, V. Croquette,   "The
elasticity of a single supercoiled DNA molecule," \emph{Science} \textbf{271}, pp. 1835-1837, 1996.


\bibitem{smith}
S. Smith, L. Finzi, C. Bustamante,   "Direct mechanical measurement
of the elasticity of single DNA molecules by using magnetic beads," \emph{Science} \textbf{258}, pp. 1122-1126, 1992.

\bibitem{busta}
C. Bustamante, J.F. Marko, E.D. Siggia, S. Smith,  "Entropic elasticity of
lambda-phage DNA," \emph{Science} \textbf{265}, pp. 1599-1601, 1994.

\bibitem{gennes}
P. -G. de Gennes,   \emph{Scaling Concepts in Polymer Physics}, Cornell University Press, Ithaca, USA, 1979.

\bibitem{odi}
T. Odijk,  "Stiff Chains and Filaments under Tension," \emph{Macromolecules} \textbf{28}, pp. 7016-7018, 1995.

\bibitem{marko2}
J.F. Marko, E.D. Siggia,   "Stretching DNA," \emph{Macromolecules} {\bf 28}, pp. 8759-8770, 1995.

\bibitem{gross}
P. Gross, N. Laurens, L.B. Oddershede, U. Bockelmann, E.J.G. Peterman, G.J.L. Wuite,  "Quantifying how DNA stretches, melts and changes twist under tension," \emph{Nat. Phys.}  \textbf{7}, pp. 731-736, 2011.

\bibitem{gore}
J. Gore, Z. Bryant, M. N\"{o}llmann, M.U. Le, N.R. Cozzarelli, C. Bustamante,  "DNA overwinds when stretched," \emph{Nature} \textbf{442}, pp. 836-839, 2006.

\bibitem{marko3}
J.F. Marko,  "Torque and dynamics of linking number relaxation in stretched supercoiled DNA," \emph{Phys. Rev. E} \textbf{76},  Article ID 021926, 2007.


\bibitem{thomas}
R. Thomas, "Recherches sur la d'enaturation des acides desoxyribonucl\'{e}iques," \emph{Biochim. Biophys. Acta} \textbf{14}, pp. 231-240, 1954.

\bibitem{doty}
S.A. Rice, P. Doty,  "The thermal denaturation of deoxyribose nucleic acid," \emph{J. Amer. Chem. Soc.} \textbf{ 79 }, pp. 3937-3947, 1957.

\bibitem{note1}
Instead, at $268$ nm, the percentage increase in absorbed light due to the opening of AT or GC \textit{bps} changes very little. See \cite{wart}.

\bibitem{li}
X.Q. Li, P. Fan, "A duplex DNA model with regular inter-base-pair hydrogen bonds", \emph{J. Theor. Biol.} \textbf{266}, pp. 374-379, 2010.

\bibitem{hwa}
T. Hwa, E. Marinari, K. Sneppen, L. Tang, "Localization of denaturation bubbles in random
DNA sequences," \emph{Proc. Natl. Acad. Sci. U.S.A.} {\bf 100}, pp. 4411-4416, 2003.

\bibitem{ares1}
S. Ares, N.K. Voulgarakis, K.{\O}. Rasmussen,  A.R. Bishop,  "Bubble Nucleation and Cooperativity in DNA Melting," \emph{Phys. Rev. Lett.} \textbf{94},  Article ID 035504, 2005.

\bibitem{ares2}
S. Ares, N.K. Voulgarakis, K.{\O}. Rasmussen,  A.R. Bishop,  "Comment", \emph{Phys. Rev. Lett.} \textbf{102},  Article ID 029602, 2009.

\bibitem{metz10}
J.H. Jeon, J. Adamcik, G. Dietler, R. Metzler, "Supercoiling Induces Denaturation Bubbles in Circular DNA," \emph{Phys.\ Rev.\ Lett.} {\bf 105},  Article ID 208101, 2010.

\bibitem{segal}
M. Bandyopadhyay, S. Gupta, D. Segal, "DNA breathing dynamics: Analytic results for distribution functions of
relevant Brownian functionals,"  \emph{Phys.\ Rev.\ E} {\bf 83},  Article ID 031905, 2011.

\bibitem{handoko}
A. Sulaiman, F.P. Zen, H. Alatas, L.T. Handoko, "Dynamics of DNA breathing in the Peyrard-Bishop model with damping and
external force," \emph{Physica D} \textbf{241}, pp. 1640-1647, 2012.

\bibitem{santa}
J. SantaLucia,   "A unified view of polymer, dumbbell, and oligonucleotide DNA
nearest-neighbor thermodynamics," \emph{Proc. Natl. Acad. Sci. USA} \textbf{95}, pp. 1460-1465, 1998.

\bibitem{owc}
R. Owczarzy, Y. You, B.G. Moreira, J.A. Manthey, L. Huang, M.A. Behlke, J.A. Walder,  "Effects of Sodium Ions on DNA Duplex Oligomers: Improved Predictions of Melting Temperatures," \emph{Biochemistry} \textbf{43}, pp. 3537-3554, 2004.

\bibitem{poland}
D. Poland, H. Scheraga,  "Occurrence of a Phase Transition in Nucleic Acid Models," \emph{J. Chem. Phys.} \textbf{45}, pp. 1464-1469, 1966.

\bibitem{fisher}
M.E. Fisher,  "Effect of Excluded Volume on Phase Transitions in Biopolymers," \emph{J. Chem. Phys.} \textbf{45},  pp. 1469-1473, 1966.


\bibitem{poland1}
D. Poland,  "Recursion Relation Generation of Probability Profiles for Specific-Sequence Macromolecules with Long-Range Correlations," \emph{Biopolymers} \textbf{13}, pp. 1859-1871, 1974.

\bibitem{jost}
D. Jost, R. Everaers,  "A Unified Poland-Scheraga Model of Oligo- and Polynucleotide DNA
Melting: Salt Effects and Predictive Power," \emph{Biophys. J.} \textbf{21}, pp. 1056-1067, 2009.

\bibitem{peliti}
Y. Kafri, D. Mukamel, L. Peliti,  "Why is the DNA Denaturation Transition First Order?", \emph{Phys. Rev. Lett.} \textbf{85}, pp. 4988-4991, 2000.

\bibitem{stella}
E. Carlon, E. Orlandini, A.L. Stella,  "Roles of Stiffness and Excluded Volume in DNA Denaturation," \emph{Phys. Rev. Lett.} {\bf 88},  Article ID 198101, 2002.

\bibitem{carlon}
R. Blossey, E. Carlon,  "Reparametrizing the loop entropy weights: effect on DNA melting curves," \emph{Phys.\ Rev.\ E} {\bf 68},  Article ID 061911, 2003.

\bibitem{hanke}
A. Hanke, M.G. Ochoa, R. Metzler,  "Denaturation Transition of Stretched DNA," \emph{Phys. Rev. Lett.} \textbf{100}, Article ID 018106, 2008.

\bibitem{seno}
D. Marenduzzo, E. Orlandini, F. Seno, A. Trovato, "Different pulling modes in DNA overstretching: A theoretical analysis,"  \emph{Phys.\ Rev.\ E} {\bf 81},  Article ID 051926, 2010.

\bibitem{pey1}
M. Peyrard, A.R. Bishop,   "Statistical mechanics of a nonlinear model for DNA denaturation," \emph{Phys. Rev. Lett.} \textbf{62}, pp. 2755-2758, 1989.

\bibitem{joy05}
M. Joyeux, S. Buyukdagli,  "Dynamical model based on finite stacking enthalpies for homogeneous and inhomogeneous DNA
thermal denaturation," \emph{Phys. Rev. E} {\bf 72},  Article ID 051902, 2005.

\bibitem{bonnet}
G. Altan-Bonnet, A. Libchaber, O. Krichevsky,  "Bubble Dynamics in Double-Stranded DNA," \emph{Phys. Rev. Lett.} \textbf{90},  Article ID 138101, 2003.

\bibitem{zocchi2}
Y. Zeng, A. Montrichok, G. Zocchi,   "Bubble Nucleation and Cooperativity in DNA Melting," \emph{J. Mol. Biol.} \textbf{339}, pp. 67-75, 2004.


\bibitem{singh}
S. Srivastava, N. Singh,  "The probability analysis of opening of DNA," \emph{J. Chem. Phys.} \textbf{134},  Article ID 115102, 2011.


\bibitem{io09}
M. Zoli,  "Path integral method for DNA denaturation," \emph{Phys.Rev. E} \textbf{79},  Article ID 041927,  2009.

\bibitem{io10}
M. Zoli,  "Denaturation patterns in heterogeneous DNA," \emph{Phys. Rev. E } \textbf{81},  Article ID 051910, 2010.

\bibitem{note2} Some works extending the PB model, including the twist around the molecule backbone, are given in, e.g. \cite{weber,barbi0}. These works however do not consider the anharmonic stacking: see Section 2.

\bibitem{engl}
S. W. Englander, N. R. Kallenbach, A. J. Heeger, J. A. Krumhansl,
A. Litwin, "Nature of the open state in long polynucleotide double helices:
Possibility of soliton excitations," \emph{Proc. Natl. Acad. Sci. U.S.A.} \textbf{77}, pp. 7222-7226, 1980.



\bibitem{sale}
M. Salerno, "Discrete model for DNA-promoter dynamics," \emph{Phys.\ Rev.\ A} {\bf 44}, pp. 5292-5297, 1991.

\bibitem{yakus}
L.V. Yakushevich, "Nonlinear DNA dynamics: hierarchy of the models," \emph{Physica D} \textbf{79}, pp. 77-86, 1994.

\bibitem{dani}
M. Daniel, M. Vanitha, "Bubble solitons in an inhomogeneous, helical DNA molecular chain with flexible strands," \emph{Phys.\ Rev.\ E} \textbf{84},  Article ID 031928, 2011.



\bibitem{proh}
E.W. Prohofsky,   "Solitons hiding in DNA and their possible significance in RNA transcription," \emph{Phys.\ Rev.\ A} {\bf 38}, pp. 1538-1541, 1988.

\bibitem{krum}
J.A. Krumhansl, J.R. Schrieffer,   "Dynamics and statistical mechanics of a one-dimensional model Hamiltonian for structural phase transitions,"   \emph{Phys. Rev. B} \textbf{11}, pp. 3535-3545, 1975.


\bibitem{morse}
P.M. Morse,  "Diatomic molecules according to the wave mechanics. II. Vibrational levels," \emph{Phys. Rev.} \textbf{34}, pp. 57-64, 1929.

\bibitem{hove}
L. van Hove, "Sur L'int\'{e}grale de Configuration Pour Les Syst\`{e}mes De Particules \`{A} Une Dimension," \emph{Physica} {\bf 16}, pp. 137-143, 1950.

\bibitem{landau}
L.D. Landau, E.M. Lifshitz,   {\it Statistical Physics}, Pergamon Press, Oxford, UK, 1980.


\bibitem{azbel1}
M.Y. Azbel,  "Generalized one-dimensional Ising model for polymer thermodynamics," \emph{J. Chem. Phys.} {\bf 62}, pp. 3635-3641, 1975.

\bibitem{pey2}
T. Dauxois, M. Peyrard, A.R. Bishop,   "Entropy-driven DNA denaturation," \emph{Phys. Rev. E} \textbf{47},  R44-47, 1993.

\bibitem{cule}
D. Cule, T. Hwa, "Denaturation of Heterogeneous DNA," \emph{Phys.\ Rev.\ Lett.} {\bf 79}, pp. 2375-2378, 1997.


\bibitem{blake}
R.D. Blake, S.G. Delcourt,  "Thermal stability of DNA," \emph{Nucleic Acids Res.} {\bf 26}, pp. 3323-3332, 1998.

\bibitem{eijck}
L. van Eijck, F. Merzel, S. Rols, J. Ollivier, V.T. Forsyth, M.R. Johnson,  "Direct Determination of the Base-Pair Force Constant of DNA from the
Acoustic Phonon Dispersion of the Double Helix", \textit{Phys. Rev. Lett.} \textbf{ 107},   Article ID 088102, 2011.


\bibitem{santos}
J.M. Romero-Enrique, F. de los Santos, M.A. Mu\~{n}oz,  "Renormalisation group determination of the order of the DNA denaturation transition," \emph{Europhys. Lett.} {\bf 89},  Article ID 40011, 2010.

\bibitem{druk}
K. Drukker, G. Wu, G.C. Schatz, "Model simulations of DNA denaturation dynamics," \emph{J. Chem. Phys.} \textbf{114}, pp. 579-590, 2001.

\bibitem{wart}
R.M. Wartell, A.S. Benight,  "Thermal denaturation of DNA molecules: A comparison of theory with experiment," \emph{Phys. Rep.} \textbf{126}, pp. 67-107, 1985.

\bibitem{benham}
R.M. Fye,  C.J. Benham, "Exact method for numerically analyzing a model of local denaturation
in superhelically stressed DNA," \emph{Phys. Rev. E} {\bf 59}, pp. 3408-3426,  1999.

\bibitem{benham1}
C.J. Benham, R.R.P. Singh, Comment on "Can One Predict DNA Transcription
Start Sites by Studying Bubbles?" \emph{Phys. Rev. Lett.} \textbf{97},  Article ID 059801, 2006.

\bibitem{io11}
M. Zoli,  "Thermodynamics of twisted DNA with solvent interaction,"  \emph{J. Chem. Phys.} \textbf{135},  Article ID 115101, 2011.

\bibitem{weber}
G. Weber, "Sharp DNA denaturation due to solvent interaction," \emph{Europhys. Lett.} \textbf{73}, pp. 806-811, 2006.

\bibitem{zdrav}
S. Zdravkovi\'{c}, M.V. Satari\'{c}, "Single-molecule unzippering experiments on DNA and Peyrard-Bishop-Dauxois model,"
\emph{Phys. Rev. E} {\bf 73},  Article ID 021905, 2006.

\bibitem{heslot}
U. Bockelmann, B. Essevaz-Roulet, F. Heslot, "Molecular Stick-Slip Motion Revealed by Opening DNA with Piconewton Forces," \emph{Phys. Rev. Lett.} \textbf{79}, pp. 4489-4492, 1997.


\bibitem{zhang}
Y.L. Zhang, W.M. Zheng, J.X. Liu, Y.Z. Chen,    "Theory of DNA melting based on the Peyrard-Bishop model," \emph{{Phys. Rev. E}} \textbf{56}, pp. 7100-7115, 1997.

\bibitem{campa}
A. Campa, A. Giansanti,  "Experimental tests of the Peyrard-Bishop model applied to the melting of very
short DNA chains," \emph{ Phys. Rev. E} \textbf{58}, pp. 3585-3588, 1998.

\bibitem{barbi0}
M. Barbi, S. Cocco, M. Peyrard,   "Helicoidal model for DNA opening," \emph{Phys. Lett. A} \textbf{253}, pp. 358-369, 1999.

\bibitem{io11a}
M. Zoli,  "Stacking interactions in denaturation of DNA fragments," \emph{Eur. Phys. J. E } {\bf 34},  Article ID 68, 2011.

\bibitem{weber1}
G. Weber, "Mesoscopic model parametrization of hydrogen
bonds and stacking interactions of RNA from melting temperatures," \emph{Nucl. Acids Res.} \textbf{41}, Article ID e30, 2013.

\bibitem{collins}
F. Zhang, M.A. Collins,   "Model simulations of DNA dynamics," \emph{Phys. Rev. E} {\bf 52}, pp. 4217-4224, 1995.

\bibitem{singh1}
A. Singh, N. Singh,  "Phase diagram of mechanically stretched DNA: The salt
effect," \emph{Physica A} \textbf{392}, pp. 2052-2059, 2013.

\bibitem{pablo}
T.A. Knotts, N. Rathore, D.C. Schwartz,  J.J. de Pablo, "A coarse grain model for DNA," \emph{J. Chem.
Phys.} \textbf{126},  Article ID 084901, 2007.

\bibitem{large}
A.M. Ababneh, Z.Q. Ababneh, C.C. Large, "DNA A-tracts bending: Polarization effects on electrostatic interactions across their minor groove", \textit{J. Theor. Biol. } \textbf{252}, pp. 742-749, 2008. 

\bibitem{io12}
M. Zoli,  "Anharmonic stacking in supercoiled DNA," \emph{J. Phys.: Condens. Matter} {\bf 24},  Article ID 195103, 2012.

\bibitem{depew}
R.E. Depew, J.C. Wang, "Conformational fluctuations of DNA helix," \emph{Proc. Natl. Acad. Sci. U.S.A.} \textbf{72}, pp. 4275-4279, 1975.

\bibitem{duguet}
M. Duguet, "The helical repeat of DNA at high temperature," \emph{Nucleic Acids Res.} \textbf{21}, pp. 463-468, 1993.

\bibitem{feyn}
R.P. Feynman,  "Space-time approach to non-relativistic quantum mechanics," \emph{ Rev.\ Mod.\ Phys.} {\bf 20}, pp. 367-87, 1948.

\bibitem{feynman}
R.P. Feynman, R. Leighton, M. Sands,  \emph{The Feynman Lectures on Physics. Vol.2}, San Francisco, Addison Wesley, USA, 2006.


\bibitem{io05a}
M. Zoli,  "Path Integral of the Two Dimensional Su-Schrieffer-Heeger Model," \emph{Phys. Rev. B} {\bf 71},  Article ID 205111, 2005.

\bibitem{io05b}
M. Zoli, "Path integral of the Holstein model with a $\phi^4$ on-site potential," \emph{Phys. Rev. B} {\bf 71},  Article ID 214302, 2005.


\bibitem{palmeri}
J. Palmeri, M. Manghi, N. Destainville, "Thermal denaturation of fluctuating finite DNA chains:
The role of bending rigidity in bubble nucleation," \emph{Phys. Rev. E} \textbf{77},  Article ID 011913, 2008.

\bibitem{joy07}
S. Buyukdagli, M. Joyeux,  "Theoretical investigation of finite size effects at DNA melting," \emph{Phys.\ Rev.\ E} {\bf 76},  Article ID 021917, 2007.

\bibitem{choi}
C.H. Choi, G. Kalosakas, K.{\O}. Rasmussen, M. Hiromura,
A.R. Bishop, A. Usheva,  "DNA dynamically directs its own transcription
initiation," \emph{Nucl. Acids Res.} \textbf{ 32}, pp. 1584-1590, 2004.

\bibitem{rapti}
Z. Rapti, A. Smerzi, K.{\O}. Rasmussen, A.R. Bishop, C.H. Choi, A. Usheva, "Healing length and bubble formation in DNA,"
\emph{Phys. Rev. E} \textbf{73},  Article ID 051902, 2006.

\bibitem{yakov}
P. Yakovchuk, E. Protozanova, M.D. Frank-Kamenetskii,  "Base-stacking and base-pairing contributions into
thermal stability of the DNA double helix," \emph{Nucleic Acids Res.} \textbf{34}, pp. 564-574, 2006.

\bibitem{metz06}
T. Ambj\"{o}rnsson, S.K. Banik, O. Krichevsky, R. Metzler,  "Sequence Sensitivity of Breathing Dynamics in Heteropolymer DNA," \emph{Phys. Rev. Lett.} \textbf{97},  Article ID 128105, 2006.

\bibitem{mazza}
C. Mazza, Strand separation in negatively supercoiled DNA, \textit{J. Math. Biol.} \textbf{51}, pp. 198-216, 2005.

\bibitem{bates}
A.D. Bates, A. Maxwell,  \emph{DNA Topology}, Oxford University Press, Oxford, UK, 2005.

\bibitem{doty1}
J. Marmur, P. Doty,  "Determination of the base composition of deoxyribonucleic
acid from its thermal denaturation temperature," \emph{ J. Mol. Biol.} \textbf{5}, pp. 109-118,  1962.

\bibitem{joy09}
M. Joyeux, A.M. Florescu,  "Dynamical versus statistical mesoscopic
models for DNA denaturation," \emph{J. Phys.: Condens. Matter} {\bf 21}, Article ID  034101,  2009.


\bibitem{bresl}
T.V. Chalikian, J. V\"{o}lker, G.E. Plum, K.J. Breslauer, "A more unified picture for the thermodynamics of nucleic acid duplex melting: A characterization by calorimetric and volumetric techniques," \emph{Proc. Natl. Acad. Sci. USA} \textbf{96}, pp. 7853-7858, 1999.

\bibitem{yera1}
E. Yeramian, "Genes and the physics of the DNA double-helix," \textit{Gene} \textbf{255}, pp. 139-150, 2000.


\bibitem{cooper}
V.R. Cooper, T. Thonhauser, A. Puzder, E. Schr\"{o}der,
B.I. Lundqvist, D.C. Langreth, "Stacking Interactions and the Twist of DNA," \emph{J. Am. Chem. Soc.} \textbf{130}, pp. 1304-1308, 2008.

\bibitem{palm13}
A.K. Dasanna, N. Destainville, J. Palmeri, M. Manghi, "Slow closure of denaturation bubbles in DNA: Twist matters", \textit{Phys. Rev. E }  \textbf{87}, Article ID 052703, 2013.

\bibitem{io13}
M. Zoli,  "Helix untwisting and bubble formation in circular DNA,"  \emph{J. Chem. Phys.} \textbf{138},  Article ID 205103, 2013.

\bibitem{dutta}
Y. Shibata, P. Kumar, R. Layer, S. Willcox, J.R. Gagan, J.D. Griffith, A. Dutta, "Extrachromosomal MicroDNAs
and Chromosomal Microdeletions in Normal Tissues", \emph{Science} {\bf 336}, pp. 82-86, 2012.



\end{thebibliography}
\end{document}